\documentclass[aip,amsmath,amssymb,jcp]{revtex4-1}
\usepackage{graphicx}
\usepackage{psfrag}
\usepackage{ae}
\usepackage{amsmath,amssymb}
\usepackage[usenames]{color}

\begin{document}

\title[]{Water diffusion in carbon nanotubes under directional electric fields: Coupling between mobility and hydrogen bonding}


\author{D\'ebora N. de Freitas}
\email{dnfreitas.naza@gmail.com}
\affiliation{Departamento de F\'isica, Universidade Federal de Ouro Preto, 35400-000 Ouro Preto, Brazil}
\affiliation{Departamento de F\'isica, Universidade Federal de Juiz de Fora, 36036-900 Juiz de Fora, Brazil}

\author{Bruno H. S. Mendon\c{c}a}
\affiliation{Departamento de F\'isica, Universidade Federal de Ouro Preto, 35400-000 Ouro Preto, Brazil}
\affiliation{Instituto de F\'isica, Universidade Federal do Rio Grande do Sul, 91501-970 Porto Alegre, Brazil}

\author{Mateus H. K\"ohler}
\email{mateus.kohler@ufsm.br}
\affiliation{Departamento de F\'isica, Universidade Federal de Santa Maria, 97105-900 Santa Maria, Brazil}

\author{Marcia C. Barbosa}
\affiliation{Instituto de F\'isica, Universidade Federal do Rio Grande do Sul, 91501-970 Porto Alegre, Brazil}

\author{Matheus J. S. Matos}
\author{Ronaldo J. C. Batista}
\author{Alan B. de Oliveira}
\email{deoliveira.alanbarros@gmail.com}
\affiliation{Departamento de F\'isica, Universidade Federal de Ouro Preto, 35400-000 Ouro Preto, Brazil}

\begin{abstract}

We have investigated the diffusion and structure of TIP4P/2005 water confined in carbon nanotubes
subjected to external electric fields.
A wide range of diameters has been used to show a highly size-dependent behavior of the water diffusion.
We also found that the diffusion is extremely affected by the intensity of the applied field.
However, is the relative direction between the field and the tube axis that causes the most intriguing behavior.
Electric fields forming angles of $0^{\circ}$ and $45^{\circ}$ with the tube axis
were found to slow down the water dynamics by increasing organization,
while fields perpendicular to the tube axis can enhance water diffusion in some cases by decreasing the hydrogen bond formation.
Remarkably, for the 1.2 nm diameter long (9,9) nanotube,
the field along the tube axis melts the water structure increasing the water mobility.
These results points out that the structure and dynamics of confined water
are extremely sensitive to external fields and suggest the use of electric fields as a facilitator for filtration processes.
\end{abstract}


\maketitle

\section{\label{sec:level1}{Introduction}}

The characterization of the molecular transport phenomena is relevant for the study of pure liquids,
mixtures and complex systems such as gels, liquid crystals and polymers.
For simple liquids, the diffusion coefficient, $D$, usually increases with temperature, $T$.
However, this is not the case of water.
The mobility of water increases with density and has a maximum~\cite{angell-jcp76}. 
This unexpected behavior is followed by a structural anomaly.
While most materials become more structured as density increases, 
under the same conditions water exhibits a maximum in their structural order parameter~\cite{errington-nature2001}.
As the order of the system increases, the structural, kinetic and thermodynamic anomalies take place~\cite{netz-jcp2001,netz-bjp2004}.

The anomalous bulk mobility of water is strongly related to their hydrogen bond (HB) network and
becomes even more anomalous under nanoconfinement.
The bonding between water molecules can be frustrated in such extreme environments~\cite{gordillo-cpl2000,gordillo-cpl2001},
and the water-surface interaction rises as a decisive factor,
often determining the fluid structural and dynamical behavior~\cite{kohler-ces2019,kohler-jpcc2019}.
For example,
inside carbon nanotubes (CNTs) water flows four to five orders of magnitude faster than would be predicted
from conventional hydrodynamic theory~\cite{majumder2005nanoscale}.
The high flux depends on the nature of the confining medium~\cite{bonthuis-jp2011},
the number of water layers~\cite{wang-langmuir2018,sahu-jcp2015},
structural transitions~\cite{kotsalis-ijmf2004,shiomi-jpcc2007,kohler-pccp2017},
and local viscosity~\cite{calabro-mrsbull2017},
to name just a few of the parameters affecting the flow enhancement factor.
For CNTs with diameter smaller than 1.2 nm,
the diffusion mechanism appears in two stages~\cite{striolo-nl2006,farimani-jpcb2011,kohler-physa2018}
reflecting the competition between wall repulsion and HB formation~\cite{wang-jcp2012,zheng-pccp2012}.

Over the past decade,
we have witnessed several attempts with different strategies
to pump water through nanotubes~\cite{bonthuis-jp2011,farimani-sr2016,kohler-ces2019}.
One alternative lies in the application of external electric fields.
For instance,
in small capillaries the electrostatic boundary conditions strongly influences the permeability of charged species.
The electrostatics of a small channel embedded in a medium of very low dielectric constant
leads to a large self-energy barrier for ions to enter the confined region~\cite{bonthuis-prl2006}.
In many biological channels,
transport of ions is facilitated by inclusion of fixed charges in the channel.
In CNTs, fixed charges outside the tube are found to affect the pressure-driven passage of water~\cite{li-pnas2007}.
Additionally, CNTs filled with water exhibit electro-osmotic flow when an electric field is applied~\cite{he-sr2014}.

The behavior of water molecules confined in CNTs subjected to external electric fields is an interesting topic by itself.
Free energy studies have demonstrated that electric fields induce water orientation
at the point that it favors the filling of small CNTs~\cite{vaitheeswaran-jcp2004}.
Theoretical and numerical studies also demonstrated a relation between a blocking temperature
(required for water to escape from the nanotube)
and a frictional energy barrier with the dynamic escape behavior of water under external electric fields~\cite{li-jpcc2016}.
Charge magnitudes between zero to 1.0 $e$/atom were applied along different lengths of nanotubes
to show that the electric charging can be used to manipulate
the resistance along the nanotube and the resistance to flow at its entrance~\cite{abbasi-jml2016},
two parameters affecting the flow rate.
Remarkably, the water dipole orientation inside a CNT can be well-tuned by the electric field,
which results in a pumping effect~\cite{su-acsnano2011}.
The dipole orientation also makes water's dielectric constant
to be lower in the interface than in the center of the tube~\cite{fumagalli-sci2018}.
Phase transitions of water molecules inside 1.2-1.3 nm diameter CNTs
induced by electric fields have been also reported~\cite{qian-jcp2014}.
In this scenario, the structure of the confined water was found to be strongly dependent on the field strength.
Recently, Winarto et al.~\cite{winarto-jcp2015}
demonstrated that electric fields can cause the formation of solid ice-like structures inside CNTs.
It is not clear, however, how the electric field affects the different
structures and layers produced when water is confined by nanotubes of different diameters.

Another open aspect demanding further discussions
is the relationship between HB formation and the dynamics of water-CNT subjected to an electric field.
Molecular Dynamics (MD) simulations of water permeation through CNTs have shown a breakage of the HB network
as the electrical interference frequency approaches to the inherent resonant frequency of HBs~\cite{kou-nl2014}.
At this point the authors found a maximum in the water net flux.
The disruption of water's HB network inside nanotubes was also pointed as the main reason
behind the high flux found when a vibrational charge was approached to the nanochannel~\cite{kou-ac2015}.
Additionally, Zhou et al.~\cite{zhou-cpl2017} showed that the motion mechanism (characterizing the diffusion)
of water confined in CNTs is changed from Fickian to ballistic and single-file diffusion
under the application of pulsed electric fields.
Investigations that can bring insights at the molecular level are more than welcome
to shade light on the mechanisms connecting the mobility with the structure and HB network of water inside CNTs
under the application of electric fields.

In this contribution,
we perform MD simulations
to study the diffusion and arrangement of water molecules inside CNTs
under external electrical fields with different intensities.
In order to account for orientation effects,
the field has been applied parallel,
transverse and perpendicular to the tube axis.
The number of HBs and the radial structuration of the water molecules has been also evaluated.
Our results show great dependence of the water diffusion on both the electrical field intensity and direction.
Besides, the HB network was found to be closely related to the diffusional behavior of water.
The remainder of the paper is organized as follows.
In the next section, the computational details and methods are described.
In section III, the main results on the confined water properties are discussed.
Summary and Conclusions are presented in the last section.

\section{\label{sec:level2}{Model and Methods}}

We used the LAMMPS package~\cite{plimpton-jcp1995} to perform MD simulations of water confined in CNTs under electric fields.
A uniform field was applied in three different directions with different intensities.
We considered ($n,n$) CNTs, with $n$ = 7, 9, 12, 16, 20 and 40.
We used the rigid four-point TIP4P/2005~\cite{abascal-jcp2005}
water model due to its good agreement with experimental results\cite{ve11,hijes-jcp2018}.
Additionally, this model has been suggested for computational studies of water flux through nanopores~\cite{prasad-pccp2018}.
We obtained a bulk diffusion coefficient of 2.3x$10^{-9}$ m$^{2}$/s,
in good agreement with theoretical~\cite{liu-jcp2016} and experiments~\cite{krynicki1978}.
The interaction potential is described by Lennard-Jones and Coulomb terms, namely:
\begin{equation}
\label{model}
U_{\alpha\beta}(r) = 4\epsilon_{\alpha \beta} 
\left[\left(\frac {\sigma_{\alpha \beta}}{r}\right)^{12}-\left (\frac{\sigma_{\alpha \beta}}{r}\right)^6\right] + \frac{1}{4\pi\epsilon_0}\frac{q_{\alpha} q_{\beta}}{r} ,
\end{equation}

\noindent where $\alpha$ and $\beta$ represents the oxygen, hydrogen or the fictitious atom $M$.
The equilibrium distance between oxygen and hydrogen atoms is $0.09572$ nm
and each hydrogen carries a positive charge $q=0.5564$,
the oxygen carries no charge.
The equilibrium distance between oxygen and $M$ is $0.01546$ nm in the direction between the two hydrogen.
The negative charge, $-1.1128$, which neutralizes the molecule is located at $M$.

\begin{table}[t!]
\begin{center}
\caption{Nanotube chirality, diameter ($d$), length (L$_{z}$) and the average amount of water molecules inside each nanotube.}
\begin{tabular}{c c c c}
\hline \hline
CNT     $\quad$ & $\quad$ d ($nm$) & $\quad$ L$_{z}$ ($nm$) & $\quad$ H$_{2}$O \\ \hline
(7,7)   $\quad$ & $\quad$ 0.957    & $\quad$ 123.465        & $\quad$ 901      \\
(9,9)   $\quad$ & $\quad$ 1.22     & $\quad$ 50.66          & $\quad$ 908      \\
(12,12) $\quad$ & $\quad$ 1.63     & $\quad$ 22.63          & $\quad$ 901      \\
(16,16) $\quad$ & $\quad$ 2.17     & $\quad$ 11.07          & $\quad$ 911      \\
(20,20) $\quad$ & $\quad$ 2.71     & $\quad$ 10.33          & $\quad$ 1440     \\
(40,40) $\quad$ & $\quad$ 5.425    & $\quad$ 7.87           & $\quad$ 5221     \\
\hline \hline
\end{tabular}
\label{t1}
\end{center}
\end{table}

For the water-CNT interactions, we used O-C Lennard-Jones parameters
defined in the work of K\"ohler and da Silva~\cite{kohler-cpl2016}:
$\varepsilon_{CO}=0.123$ kcal/mol and $\sigma_{CO}=0.326$ nm.
The Lennard-Jones and Coulomb interactions cutoff was set to $1$ nm.
In all simulations the geometry of the water molecules was constrained by the SHAKE algorithm~\cite{ryckaert-jcp1977}.
Long-range Coulomb interactions was handled using the {\it Particle-Particle Particle-Mesh} (PPPM) method.

The simulations were conducted as follows.
Initially, the nanotubes were connected to two water reservoirs with pressure kept at $1$ atm
by the Parrinello-Rahman barostat~\cite{parrinello-jap1981}
and at temperature T$=300$ K controlled by Nos\'e-Hoover thermostat~\cite{nose-mp1984}.
The system was equilibrated for $10$ ns with timestep set as $1$ fs.
After system equilibration, the average number of water molecules within each nanotube were collected.
The amount of water molecules along with the nanotube details are given in Table~\ref{t1}.

\begin{figure}[t!]
\begin{center}
\includegraphics[width=11cm]{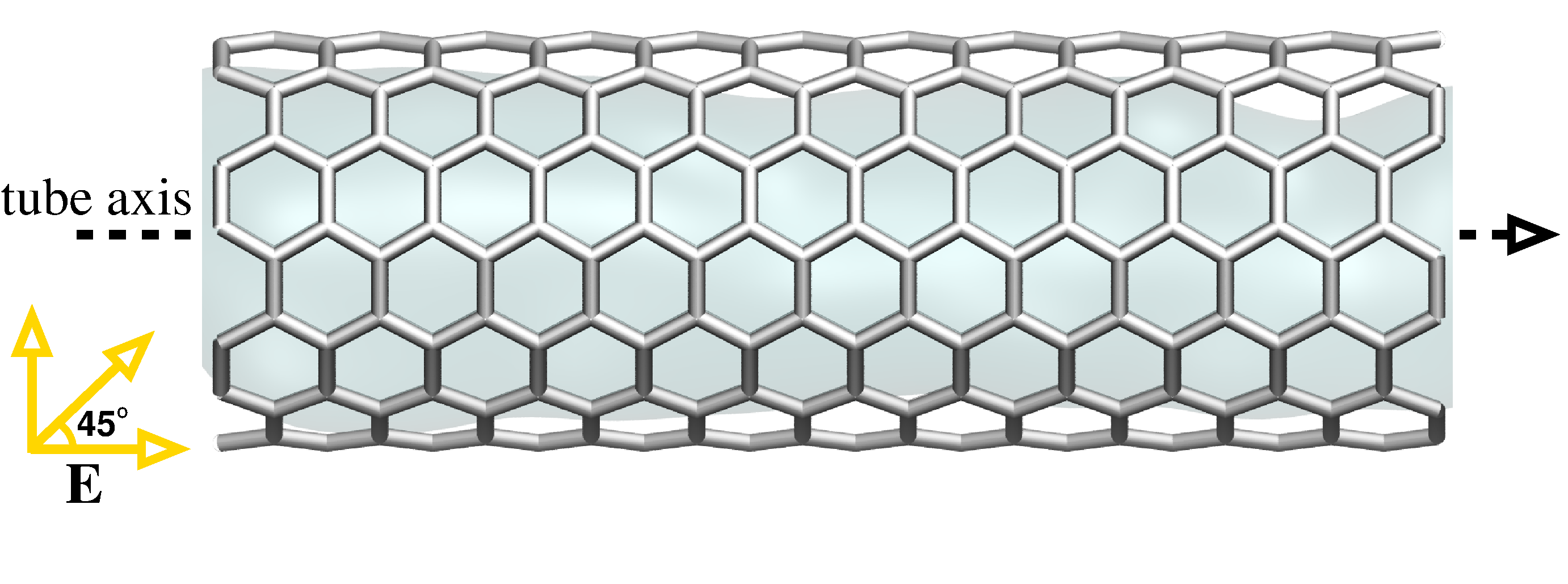}
\end{center}
\caption{Schematic depiction of the computational framework:
side view of a CNT (7,7) filled with water
under an electric field $E$.
The field can be either parallel ($0^{\circ}$), perpendicular ($90^{\circ}$) or forming an angle of $45^{\circ}$ to 
the nanotube axis.}
\label{fig:fig01}
\end{figure}

Then, the reservoirs were removed and periodic boundary conditions applied in the axial direction of the tubes,
as shown in Figure~\ref{fig:fig01}.
It results in isolated infinite nanotubes.
Finally, simulations of $10$ ns of data accumulation are performed in the canonical NVT ensemble.

We analyzed the diffusion mechanism of a fluid by the scaling behavior between the mean squared displacement (MSD) and 
time~\cite{farimani-jpcb2011}:
\begin{equation}
\langle{|\vec{r}(t)-\vec{r}(0)|^{2}}\rangle=ADt^n
\end{equation}
\noindent where the angular brackets denote an average over time origins and all water molecules, $\vec{r}\left ( t \right )$ is the 
displacement of a molecule during the time interval $t$, $A$ is a constant equal to $1$ and $D$ stands for the diffusion coefficient. 
The $n$ exponent refers to the diffusion regime: {\it{n}} = 1 for the linear  Fickian diffusion, {\it{n}} $>$ 1 for supperdiffusive 
and {\it{n}} $<$ 1 for subdiffusive regime. The statistical error in the diffusion measurements could  be reduced by averaging 
over all the MSD components. The nanopore confinement in the $x$ and $y$ directions hinders the radial displacement of the 
molecules. Therefore, the radial diffusion is almost zero for all cases studied here and only the axial diffusion D$_z$ will 
be considered.

\section{\label{sec:level3}{Results and discussion}}

\begin{figure}[t!]
\includegraphics[width=12cm]{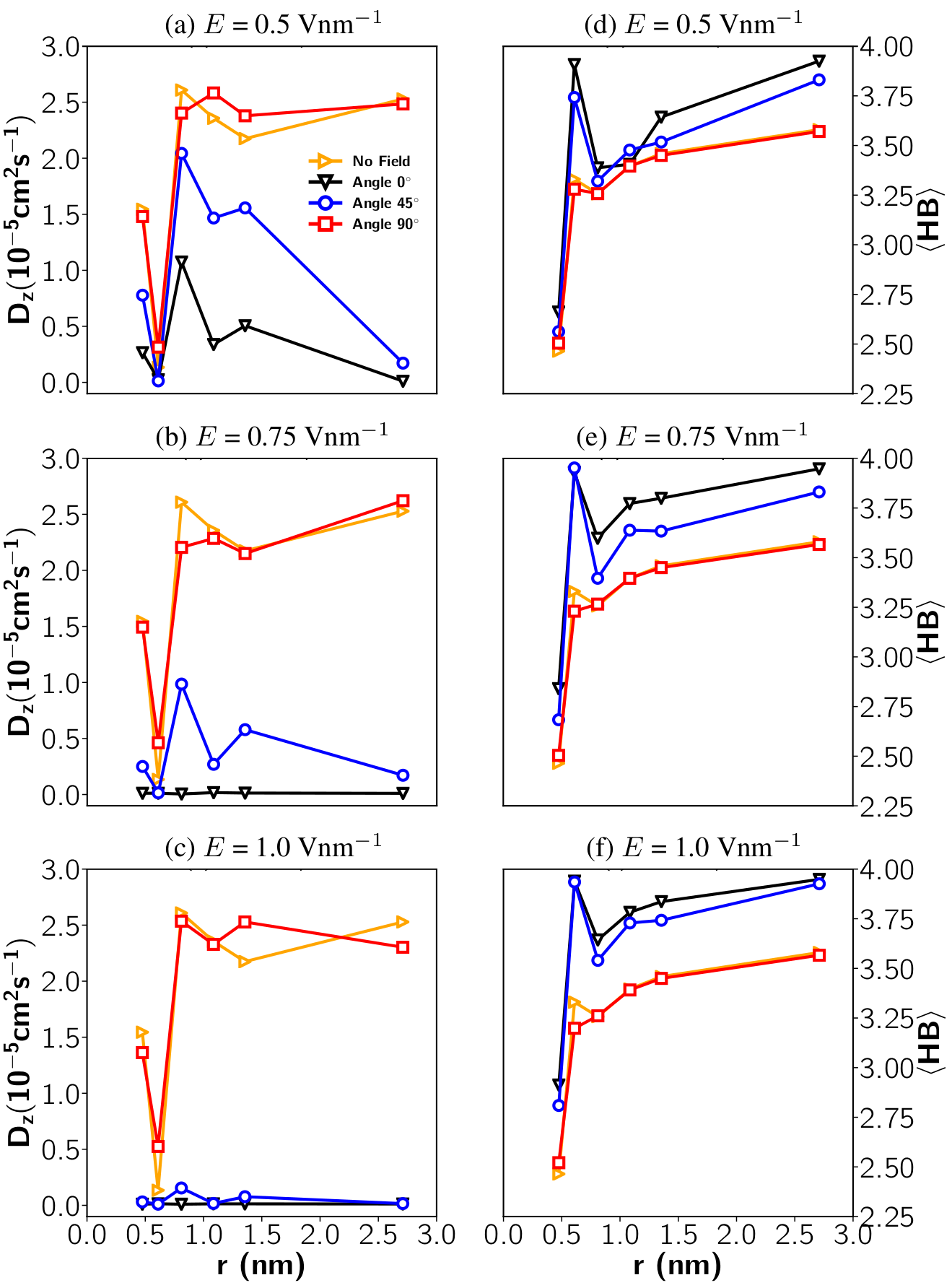}
\caption{Water diffusion coefficient (left) and number of HBs (right)
as a function of the nanotube radius $r$ under external electric fields
$E=$ (a, d)  0.5 V/nm, (b, e) 0.75 V/nm, and (c, f) 1.0 V/nm applied in different directions.}
\label{fig:fig02}
\end{figure}

An external electric field can strongly influence the behavior of fluids in nanoenvironments~\cite{kohler-ces2019}.
As a polar substance water is not only susceptible to the electrical field magnitude,
but also to the field direction inside the channel.
Consequently, if one wishes to control water flux inside nanotubes,
it is important to understand the effect of electric fields on the water properties at the molecular level.

In Figure~\ref{fig:fig02}(a)-(c) we plot the diffusion coefficient as a function of the nanotube radius
for $E=$ 0.5, 0.75 and 1.0 V/nm.
As shown in Figure~\ref{fig:fig01}, the fields are applied in three different directions:
parallel to the tube axis (referred as $0^{\circ}$),
perpendicular ($90^{\circ}$) and forming an angle of $45^{\circ}$ with the tube axis.
In the case of no electric field,
the diffusion is not monotonic with the nanotube radius.
While water freezes inside 1.2 nm diameter long (9,9) nanotube,
for larger nanotubes the diffusion increases with $r$ and tends to the bulk value,
which is in accordance with the work of Farimani and Aluru~\cite{farimani-jpcb2011}.

Remarkably, when a 0.5 V/nm field is applied parallel to the tube axis ($0^{\circ}$)
the diffusion of water decreases drastically for all the tube radii,
and tends to zero for larger nanotubes,
as shown in Figure~\ref{fig:fig02}(a).
This is in accordance with previous theoretical works~\cite{winarto-jcp2015}.
We also computed the number of HBs using a geometrical criteria:
$\alpha\leq 30^{\circ}$ and $|\vec{r}_{OO}|\leq 3.50 \mbox{ \AA}$,
where $\alpha$ is the $OH\cdots O$ angle and $|\vec{r}_{OO}|$ is the distance between two oxygen atoms.
We found that the decrease in mobility goes along with an increase in the number of HBs,
as can be seen in Figure \ref{fig:fig02}(d).
In other words, the electric field applied along the tube axis increases the connection through hydrogen bonds
and consequently hinders the water mobility.
For the $45^{\circ}$ field,
we observe intermediate values between the parallel and the perpendicular case,
with a decrease in the diffusion and an increase in the number of hydrogen bonds.
When we apply the field perpendicular to the tube axis
a minute increase in the water diffusion is observed for intermediate tube radii,
especially in the CNT (9,9),
where we finally found some water mobility induced by the electric field.
Importantly, there is a local maximum in water diffusion between $r=$ 0.8 and 1.0 nm for all field directions.

If we increase the field magnitude to 0.75 V/nm, Figures~\ref{fig:fig02}(b) and (e),
we can see that as in the former case the external field influences both diffusion and the hydrogen bond network of the confined water.
However, the effect is enhanced and now we can observe zero water mobility as the field is applied in the axial direction (angle $0^{\circ}$),
regardless of the nanotube radius.
The phenomenon is again followed by an increase in the number of HBs,
as shown in Figure \ref{fig:fig02}(e).
Again, the $45^{\circ}$ field leads to a dynamics somehow in between the parallel and perpendicular case,
still strongly reducing the mobility and increasing the number of HBs in comparison to the no field situation.
On the other hand,
as we apply the electric field perpendicular to the tube axis
the diffusion coefficient features similar results as in the no field condition,
although with some particular differences.
For instance, in the CNT (9,9) the perpendicular field enhances the water mobility at the cost of breaking some HBs,
while in the CNT (12,12) the very opposite happens,
mobility is decreased and new HBs are formed in comparison with the no field case.


\begin{figure}[t!]
\includegraphics[width=11cm]{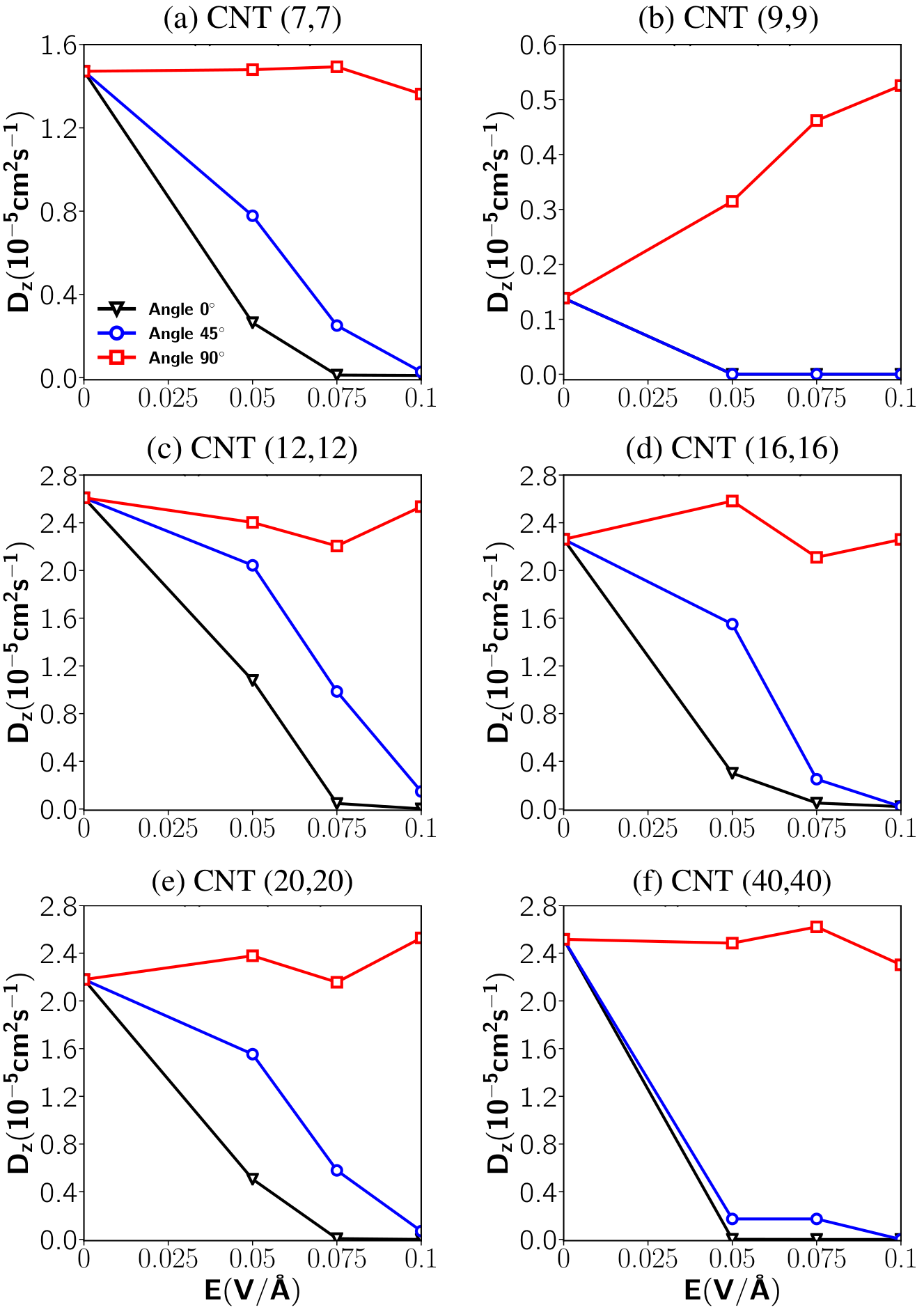}
\caption{Water diffusion coefficient as a function of the electric field's magnitude.}
\label{fig:fig03}
\end{figure}

\begin{figure}[t]
\begin{center}
\includegraphics[width=10cm]{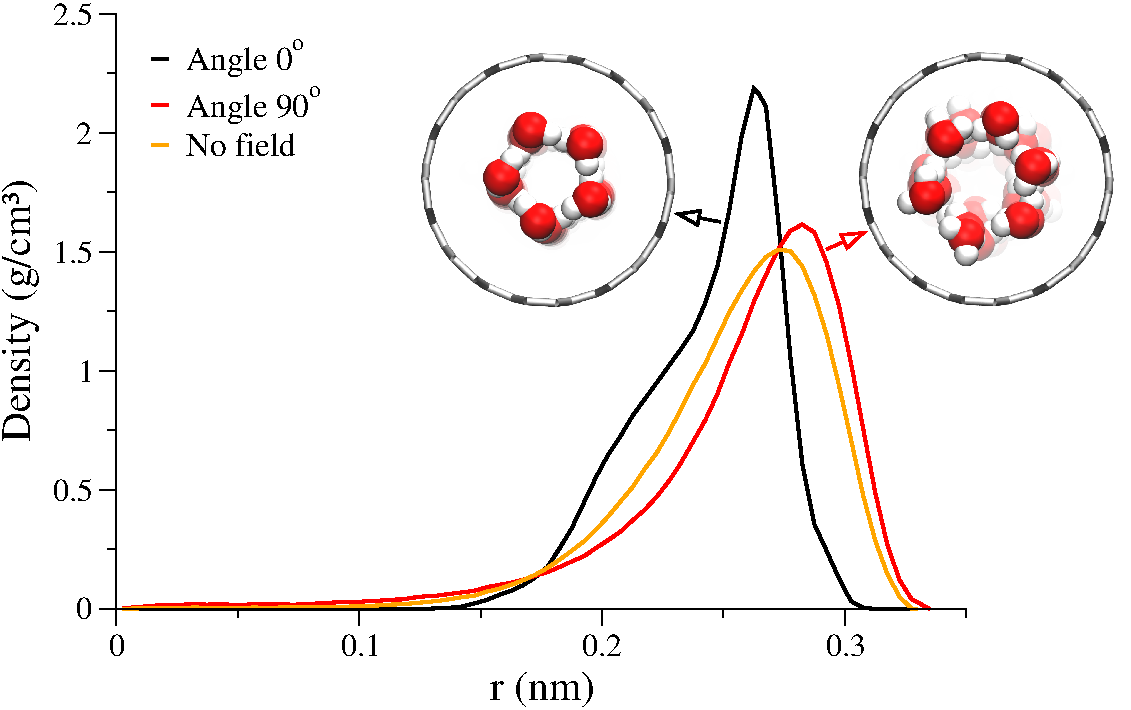}
\end{center}
\caption{Radial density profile of oxygen atoms inside CNT (9,9) for $E=$ 1.0 Vnm$^{-1}$
applied parallel (black line) and perpendicular (red line) to the tube axis.
The orange line stands for no applied field and $r=0$ is at the center of the tube.}
\label{fig:fig04}
\end{figure}

In Figure~\ref{fig:fig02}(c) we show the water diffusion
as a function of $r$ for $E=$ 1 V/nm.
At this point, the diffusion fall away for both the parallel and the $45^{\circ}$ fields,
exhibiting zero mobility (frozen) and highly enhanced HB formation, as can be seen in Figure~\ref{fig:fig02}(f).
This corroborates the idea of a coupling between the water dynamics and the HB network under external electric fields.
Furthermore, as in the former case, $E=$ 0.75 V/nm, here we can observe a strong correlation between the no field and the orthogonal cases.
Interestingly, the perpendicular field induces water inside CNTs (9,9) and (20,20) to exhibits higher diffusion than the zero field condition.

In Figure~\ref{fig:fig03} we present the water diffusion coefficient as a function of the applied field intensity for each direction.
It becomes clear that the perpendicular field induces larger water diffusion compared with that of parallel and $45^{\circ}$ case for all nanotube radii.
Surprisingly, water diffusion inside CNTs subjected to perpendicular electric fields is not significantly affected by the field intensity.
The exception comes from the water confined in CNT (9,9), where the diffusion increases almost linearly with $E$
and becomes $\sim$3 times larger when we increase the perpendicular field from 0 to 1 V/nm.
For other directions ($0^{\circ}$ and $45^{\circ}$) the water diffusion is decreased as we increase $E$
and becomes almost zero for $E=$ 1 V/nm.

An important aspect regarding the mobility of confined water
is the close relation between structural and dynamical behavior~\cite{kohler-ces2019,kohler-pccp2017,kohler-jpcc2019,pascal-pnas2011,cicero-jacs2008}.
By applying an electric field, we have observed dynamical modifications
that can be related to structural transitions of water molecules inside the tube.
For instance, Winarto et al.~\cite{winarto-jcp2015}
found that the water dipole moments can align with the electric field inside the nanotube
to increase water density and form ordered ice-like structures.
This mechanism induces a transition from liquid to ice nanotubes in a wide range of CNT diameters.
In order to clarify the influence of field direction in the water structure,
we show in Figure~\ref{fig:fig04} the radial density profile of the oxygen atoms as a function of $r$ for the CNT (9,9).
This quantity is calculated by dividing the inner of the CNT in concentric cylindrical shells
and averaging the number of oxygen atoms in each shell along the simulation. 
We can observe that the parallel electric field ($0^{\circ}$) induces water molecules
to occupy positions closer to the center of the tube as compared to the no field case.
Additionally, we observe highly ordered ring-like structures,
as shown in the inset featuring a frontal snapshot of the system.
This strongly packed structure can be linked to an ice-like phase,
in accordance with the work of Winarto et al.~\cite{winarto-jcp2015},
which can explain the lowering in water diffusion.
It is also interesting to note that as the oxygen atoms aligns with the parallel field, they are carried away from the wall.
This distance from the wall, also recognized as a ``dewetting'' transition,
can lead to a decrease in the dielectric constant as suggested in experiments with planar confinement~\cite{fumagalli-sci2018}.

On the other hand, as we apply the electric field perpendicular ($90^{\circ}$) to the tube axis,
the water molecules are pushed closer to the hydrophobic CNT wall,
breaking the ring structure and increasing the mobility, as shown in Figure~\ref{fig:fig03}(b).
The mechanism behind this enhanced diffusion can be understood in terms of the energetic stability,
since it is energetically unfavorable for water to be closer to the wall.
Particularly,
in this case the field affects the water's dipole moment melting its ordered structure at the nanotube wall,
leading now to a ``wetting'' state~\cite{kayal-jcp2015}.
As a result, the water-wall repulsion increases water's mean displacement.
In other words, in the search for a more stable structure
the water molecules experience an increment in mobility.


\section{\label{sec:level4}{Summary and Conclusions}}

We studied the influence of electric fields in diffusion, HB network and structure of water confined in CNTs through MD simulations.
We found size-dependent diffusion of water under several electric field directions and intensities.
There is always a local maxima in diffusion for intermediate nanotubes ($r=$ 0.8-1.0 nm).

We also found that large fields induce the orientation of water's dipole moment and in the case of fields applied parallel to the tube, 
the network becomes more organized and the diffusion decreases.
For fields applied perpendicular to the nanotube axis, 
water dynamics is similar with the no field case for most of the CNTs.
The exception is the CNT (9,9) where the diffusion is enhanced by a factor of 3 when we increase the perpendicular field from 0 to 1 V/nm.

These results open the possibility of tuning the water flux inside a CNT by only varying the direction of the applied electric field.
It is particularly relevant for application in nanotube membranes,
where the water's dipole moment can constitute as an important factor for flux control.

\begin{acknowledgements}
This work was financed in part by Coordena\c{c}\~ao de Aperfei\c{c}oamento de Pessoal de N\'ivel Superior -
Brasil (CAPES) - Finance Code 001.
The authors also acknowledge the Brazilian agencies CNPq and INCT-Fcx.
ABO thanks the Brazilian agency FAPEMIG for financial support through the Pesquisador Mineiro grant.
\end{acknowledgements}

\bibliographystyle{aip}
\bibliography{reference}

\end{document}